\documentclass[conference]{IEEEtran}
\IEEEoverridecommandlockouts
\usepackage{cite}
\usepackage{amsmath,amssymb,amsfonts}
\usepackage{color}
\definecolor{blue}{rgb}{0,0,1}

\usepackage{threeparttable}
\usepackage{booktabs}
\usepackage{booktabs}
\usepackage{multirow}
\usepackage{hyperref}

\usepackage{algorithmic}
\usepackage{graphicx}
\usepackage{textcomp}
\usepackage{xcolor}
\def\BibTeX{{\rm B\kern-.05em{\sc i\kern-.025em b}\kern-.08em
    T\kern-.1667em\lower.7ex\hbox{E}\kern-.125emX}}
\usepackage{booktabs}
\hypersetup{hidelinks}

\begin{document}

\title{Physics-Informed Appliance Signatures Generator for Energy Disaggregation
%
}
\makeatother

\author{\IEEEauthorblockN{Ilia Kamyshev, Sahar Moghimian Hoosh, Henni Ouerdane}
\IEEEauthorblockA{\textit{Center for Digital Engineering}\\
\textit{Skolkovo Institute of Science and Technology}\\
Moscow, Russia \\
Ilia.Kamyshev@skoltech.ru, Sahar.Moghimian@skoltech.ru, H.Ouerdane@skoltech.ru}
}

\maketitle

\begin{abstract}
Energy disaggregation is a promising solution to access detailed information on energy consumption in a household, by itemizing its total energy consumption. However, in real-world applications, overfitting remains a challenging problem for data-driven disaggregation methods. First, the available real-world datasets are biased towards the most frequently used appliances. Second, both real and synthetic publicly-available datasets are limited in number of appliances, which may not be sufficient for a disaggregation algorithm to learn complex relations among different types of appliances and their states. To address the lack of appliance data, we propose two physics-informed data generators: one for high sampling rate signals (kHz) and another for low sampling rate signals (Hz). These generators rely on prior knowledge of the physics of appliance energy consumption, and are capable of simulating a virtually unlimited number of different appliances and their corresponding signatures for any time period. Both methods involve defining a mathematical model, selecting centroids corresponding to individual appliances, sampling model parameters around each centroid, and finally substituting the obtained parameters into the mathematical model. Additionally, by using Principal Component Analysis and Kullback-Leibler divergence, we demonstrate that our methods significantly outperform the previous approaches.

\end{abstract}


\section{Introduction}
Energy disaggregation, also known as non-intrusive load monitoring (NILM), is a data-driven method to break down the total energy consumption of a household into its individual appliance-level components using a single meter \cite{192069}. The granular data provided by energy disaggregation enables end-users to discover energy-saving opportunities, identify energy vampires, detect leakage points, and malfunctions of appliances \cite{NAJAFI2018377,9713260}. To understand the importance of energy disaggregation, consider an analogy to an itemized bill from a grocery store listing the price of each item purchased. In the same fashion, disaggregation algorithms offer detailed electricity bill for households or commercial facilities that can be helpful in the identification of irrational energy consumption. At a large scale, it helps utilities implement the demand response programs and improve load forecasting accuracy \cite{10116413}.

Running data-driven models on limited data can result in higher generalization error due to the overfitting problem. Recent studies have encountered challenges with poor disaggregation accuracy on new unseen households due to a lack of sufficient labeled data \cite{article1}. Most of the known data collections contain a relatively small amount of appliances and their signatures, while real-world households or facilities typically contain dozens or even hundreds of different appliances. Moreover, there is usually a bias in number of signatures towards most frequently used appliances that causes data imbalance \cite{en12091696}.

To address these challenges, synthetic datasets are a promising solution for balancing the data and increasing its diversity \cite{IQBAL2021106921}. The number of publicly available synthetic datasets for energy disaggregation is limited to four, namely: SmartSim \cite{7778841}, Automated model builder for appliance loads (AMBAL) \cite{8340657},  Simulated high-frequency energy disaggregation (SHED) \cite{HENRIET2018268}, and  Synthetic energy dataset (SynD) \cite{article2}. 

SmartSim, the first synthetic dataset proposed for NILM, generates both the aggregate power data for simulated homes, as well as power data for each appliance inside with a sampling rate of 1 Hz over almost seven days. Their methodology for generating device models is based on both empirical and statistical methods, and it encompasses models for 25 distinct appliances. The AMBAL dataset is recorded at a sampling rate of 1 Hz for a day based on real-world power consumption data collected by smart plugs. AMBAL's approach extracts active appliance usage segments, segmenting them further by power consumption changes, and fitting every segment into two predefined basic models to find the best fit. The parametrized model with the lowest mean absolute percentage error (MAPE) value is chosen as the best fit. AMBAL contains 14 different appliances and it requires manual interaction of the user to specify the MAPE value. The SHED dataset consists of 8 commercial buildings with a total of 66 appliances. The dataset has been learned on three publicly available datasets (PLAID \cite{8379795}, COOLL \cite{picon2016cooll}, and Tracebase \cite{6388037}) and one private dataset. In this work, a data generator algorithm is proposed that models the current flowing through an electric appliance. The current of a specific device is modeled using a matrix factorization approach to decompose high-frequency current waveforms into signatures and activations. SynD is a synthetic dataset of energy usage profiles for 21 household appliances over 180 days, collected at a sampling rate of 5 Hz. The simulator selects power consumption patterns based on predefined categories and then interpolates the patterns to simulate real-world variability. It randomly selects power-on times for appliances from predefined time windows based on appliance type. The mains signal in SynD is derived by aggregating individual appliance-level power signals. Moreover, the dataset only includes single-phase appliances due to data collection cost constraints. 

While synthetic data has helped to improve the generalization of energy disaggregation algorithms, it is still limited by the number of appliances, buildings, sampling frequency, and measurement duration. All existing synthetic datasets are generated with low-frequency resolution, and this limitation in sampling rate restricts engineers' choice of disaggregation algorithms, which can result in significant deviations from the targeted problem. Additionally, the limited number of appliances in synthetic datasets may not be sufficient to spot hidden nonlinear relations among appliances of different types.

Prior to this work, we attempted to develop two physics-informed appliance data generators for high sampling rate (kHz) and low sampling rate (Hz) signals \cite{10202926}. After half a year of experiments, it turned out that both methods have significant drawbacks. Namely, their corresponding distributions are too different from the corresponding real ones. Besides, there is a lack of transparency in setting up the parameters of underlying distributions.

In this paper, we propose two novel physics-informed methods to generate appliance signatures. The first method generates signatures at a high sampling rate, while the second at a low sampling rate. Both methods have several advantages over previous works. First, our methods have transparent and intuitive control over the underlying distributions. Second, they are capable of simulating arbitrarily large numbers of appliances and their signatures. Third, they do not require input data, but rather prior knowledge of the physics of a process. Finally, our methods can also approximate actual appliances by using a proper parameterization.

The paper is organized as follows. In section \ref{sec:main}, we provide a tutorial on how to derive the proposed methods. Next, in section \ref{sec:val}, we verify the fairness of the generated data in relation to the real-world datasets. Section \ref{sec:concl} concludes the paper.
 
\section{Physics-Informed Data Generation}
\label{sec:main}

Below, we present two physics-informed appliance signatures generators for the two types of sampling rate, respectively.

\subsection{High Sampling Rate Signatures}
\label{sec:synth-hf}
\begin{figure}[b!]
  \centering
  \includegraphics[width=.5\textwidth]{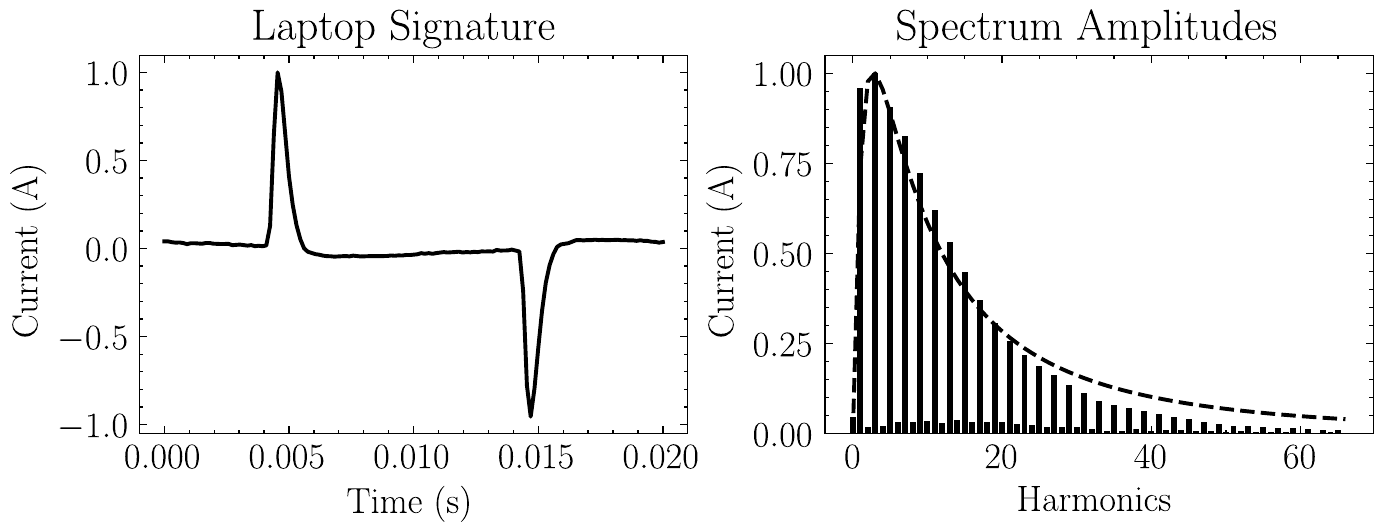}
  \caption{Single cycle of laptop's high sampling rate signature (left) and its corresponding spectrum amplitudes (right). The amplitude of both graphs is set to 1.}
  \label{fig:laptop}
\end{figure}

\begin{figure}[b!]
  \centering
  \includegraphics[width=.5\textwidth]{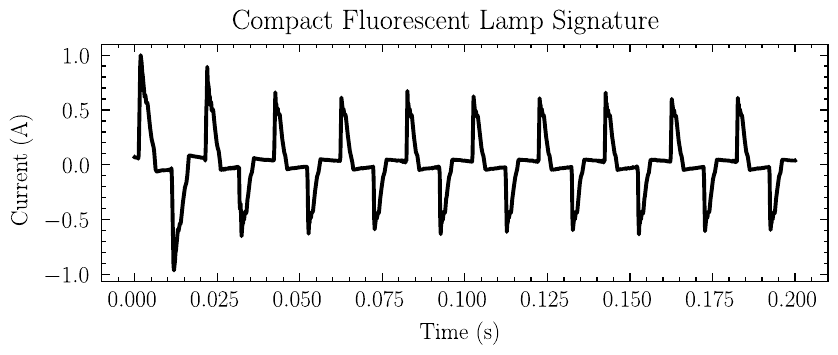}
  \caption{Ten cycles of compact fluorescent lamp's high sampling rate signature that show exponential amplitude decay and spectrum fluctuations. The amplitude is set to 1.}
  \label{fig:cfl}
\end{figure}

Prior to this work, we analysed high-resolution appliance signatures from two public datasets PLAID and WHITED \cite{inproceedings4}. It was found that there are several common features that sufficiently describe the nature of oscillatory waveforms produced by appliances:
\begin{enumerate}
    \item distribution of harmonic amplitudes follows the probability density function of log-normal distribution.
    \item presence of only odd/even or both orders of harmonics.
    \item phase shift is in the interval from $-\pi/2$ to $\pi/2$.
    \item spectrum and amplitude may vary over the time.
    \item exponential decay of a transient process.
\end{enumerate}
The illustration of some of these properties is given in Fig.~\ref{fig:laptop} and Fig.~\ref{fig:cfl}.

The spectrum of $n$ harmonics can be expressed as a set of complex variables $z=\{z_0, z_1,\ldots,z_n\}$, where $z_i=\text{Re}_i+j\cdot\text{Im}_i$ with $j^2=-1$, and the corresponding waveform is obtained using the inverse Fourier transform as $w=\mathcal{F}^{-1}[z]$. Here, we model the real and imaginary components of the complex number $z$ as follows:
\begin{equation}
\label{eq:re}
    \text{Re} = \text{Re}' + r\cdot\cos{\phi},
\end{equation}
\begin{equation}
\label{eq:im}
    \text{Im} = \text{Im}' + r\cdot\sin{\phi},
\end{equation}
where $\text{Re}'$ and $\text{Im}'$ are centroid coordinates in a complex plane that define the uniqueness of an appliance, and $r$ and $\phi$ are the radius and angle in the complex plane respectively. To incorporate the given physics, we first assert that spectrum amplitudes follow the probability density function of log-normal distribution:
\begin{equation}
\label{eq:z}
    z_i = z_i\cdot\frac{1}{i\cdot\sigma\sqrt{2\pi}}\cdot\text{exp}\left(-\frac{(\ln{i-\mu})^2}{2\sigma^2}\right),
\end{equation}
where $i>0$, $\mu$ and $\sigma$ are positive real numbers also known as shape parameters. Next, by specifying $m=$ 0 or 1 and $d=1$, we drop either the odd or the even harmonics:
\begin{equation}
\label{eq:dropout}
    z_i = 0, \text{where}\ i>1 \land d=1\land\ i\mod 2+m = 0.
\end{equation}
Setting $d=0$ will keep all the harmonics.

Further, we impose the constraint on phase shift by placing the real and imaginary components of the complex variable $z_i$ inside the third and fourth quadrants of the complex plane i.e., $\text{Im} < 0$. 

\begin{figure}[b!]
  \centering
  \includegraphics[width=.5\textwidth]{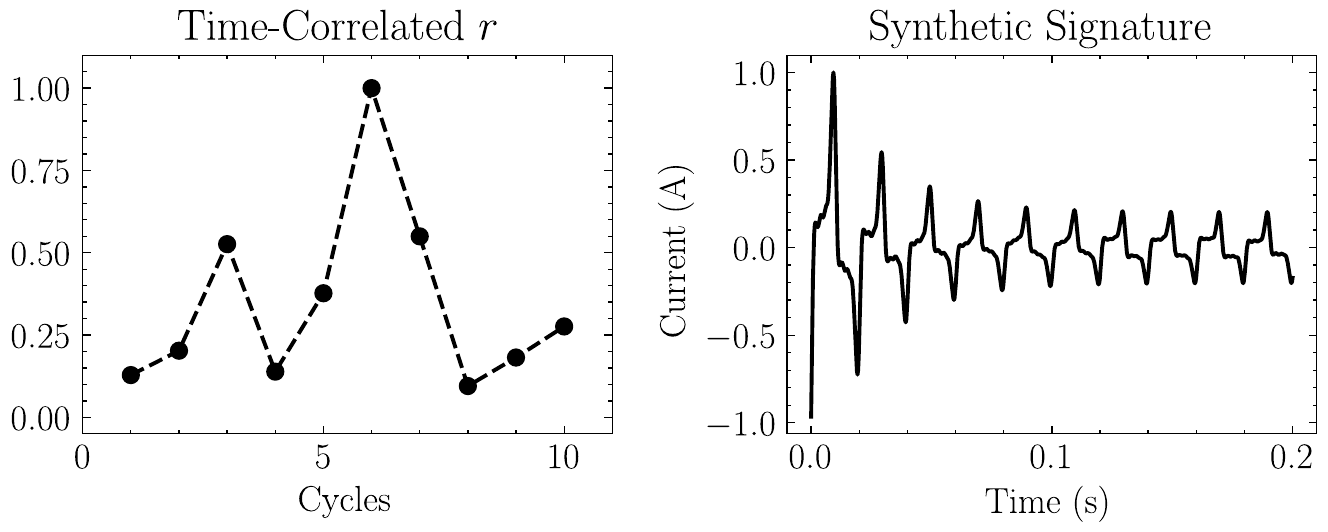}
  \caption{Time-correlated random variable $r$ (left). Ten cycles of synthetic signature (right) which spectrum is time-correlated with accordance to the graph on the left. The amplitude of both graphs is set to 1.}
  \label{fig:ar1}
\end{figure}

In the real data, the spectrum may vary over the time. To make a waveform of $p$ cycles with floating spectrum, we model $r$ and $\phi$ as autoregressive process:
\begin{equation}
\label{eq:ar1r}
    r_{t,i} = |\rho\cdot r_{t-1,i} + \epsilon_{t,i}|,
\end{equation}
\begin{equation}
\label{eq:ar1phi}
    \phi_{t,i} = |\rho\cdot\phi_{t-1,i} + \epsilon_{t,i}|\mod 2\pi,
\end{equation}
where $|\rho|<1$ is a parameter of autoregressive process, $\epsilon$ is a white noise. The example of time-correlated $r$ with $\rho=0.5$ and unit variance is given on the left diagram in Fig.~\ref{fig:ar1}.

Given the set of harmonics $z_t$ for cycle $t$, the discrete inverse Fourier transform can be used to obtain a single-cycle waveform of amplitude $a$:
\begin{equation}
\label{eq:w}
    w_t=a\cdot\frac{\mathcal{F}^{-1}[z_t]}{\max{|\mathcal{F}^{-1}[z_t]|}}.
\end{equation}
One can also model amplitude as time-correlated variable $a_t$ by using Eq.~\eqref{eq:ar1r}.

The waveform of $p$ cycles can be obtained via concatenation:
\begin{equation}
\label{eq:concat}
    w=\text{concat}(w_1, w_2,\ldots,w_p).
\end{equation}

The transient process can be modelled as in the theory of electric circuits:
\begin{equation}
\label{eq:w'}
    w' = w\cdot (1+(A-1)\cdot\text{exp}(-\tau\cdot t)),
\end{equation}
where $A$ is the peak to steady-state amplitude ratio, $\tau$ is a time constant, and $t$ is a discrete time.

The mathematical model $w'$ of an oscillatory waveform enables to simulate a wide range of appliance consumption signatures. To generate different types of appliances and their corresponding signatures, we define centroid $Z$ for each appliance $k$ as $Z_k=\{n_k, \text{Re}_k',\text{Im}_k', \mu_k, \sigma_k, m_k, d_k, \rho_k, a_k, A_k, \tau_k\}$. The distance between centroids is proportional to the similarity of appliances, and we recommend to sample all centroid coordinates from uniform distribution, except $n_k$ that can be sampled from Poisson distribution.

\begin{figure}[b!]
  \centering
  \includegraphics[width=.5\textwidth]{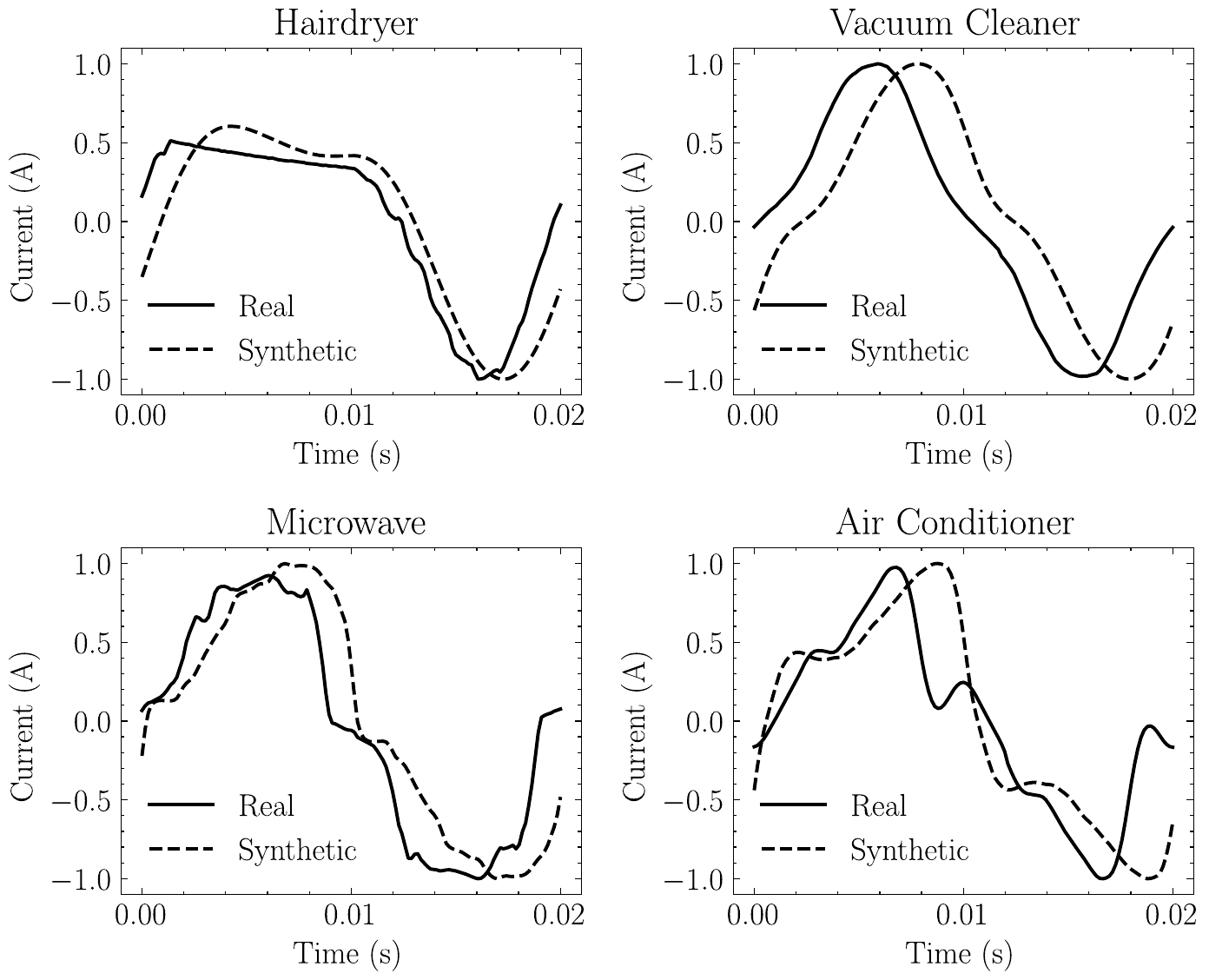}
  \caption{Four real appliances and the most similar to them synthetic appliances. The amplitude of all graphs is set to 1.}
    \label{fig:4apps}
\end{figure}

Once centroids are specified, the parameters $r, \phi$ for each appliance $k$ should be computed as in Eqs.~\eqref{eq:ar1r},~\eqref{eq:ar1phi} with the white noise $\epsilon$ of variance $\text{Var}_d$ that controls the diversity of signatures. We suggest to sample amplitudes $a$ from half-normal distribution with mean $a_k$ and variance $\text{Var}_d$. Number of cycles per signature $p$ can be set as constant for convenience. After sampling, the parameters should be substituted in Eqs.~\eqref{eq:re},~\eqref{eq:im},~\eqref{eq:z},~\eqref{eq:dropout},~\eqref{eq:w},~\eqref{eq:concat},~\eqref{eq:w'} to obtain signatures of synthethic appliances. For demonstrational purpose, using the proposed approach we simulated four synthetic appliances that parameters were chosen at random. By using cosine similarity measure, we matched four most similar real appliances with the obtained ones, they are hairdryer, vacuum cleaner, microwave and air conditioner (see Fig.~\ref{fig:4apps}). 

\subsection{Low Sampling Rate Signatures}
\label{sec:synth-lf}
Low sampling rate signatures or RMS waveforms are another type of appliance signatures. It is challenging to approximate such waveforms by continuous functions as they contain many jump-discontinuities. However, this task can be significantly simplified by dividing each appliance signature into primitive cycles. By primitive cycle, we mean a continuous interval of non-zero consumption. After inspecting the REDD \cite{REDD} and UK-DALE \cite{article5} datasets, we identified several frequently occurring primitive cycles for most appliances. We define 5 basis functions which products can approximate these cycles on the discrete interval $[0, \Delta t]$:
\begin{equation}
\label{eq:p1}
    p_1=a,
\end{equation}
\begin{equation}
\label{eq:p2}
    p_2=1+A\cdot\text{exp}(-\tau\cdot t),
\end{equation}
\begin{equation}
\label{eq:p3}
    p_3 = 1 + \mathcal{L}^{-1}\left[\frac{q_0}{q_1 \cdot s^2 + q_2\cdot s + q_3}\right],
\end{equation}
\begin{equation}
\label{eq:p4}
    p_4 \sim \mathcal{N}(\mu=1,\sigma_n^2),
\end{equation}
\begin{equation}
\label{eq:p5}
    p_5 \sim \text{Beta}(\alpha,\beta),
\end{equation}
where $a$ is the amplitude, $A$ is the peak to steady-state amplitude ratio, $\tau$ is a time constant, $q_0$, $q_1$, $q_2$, $q_3$ are transfer function parameters, $\mathcal{L}^{-1}$ is the inverse Laplace transform, $\sigma_n^2$ is a noise variance, and $\text{Beta}(\alpha,\beta)$ is the beta distribution with shape parameters $\alpha$ and $\beta$. 

\begin{figure}[b!]
  \centering
  \includegraphics[width=.5\textwidth]{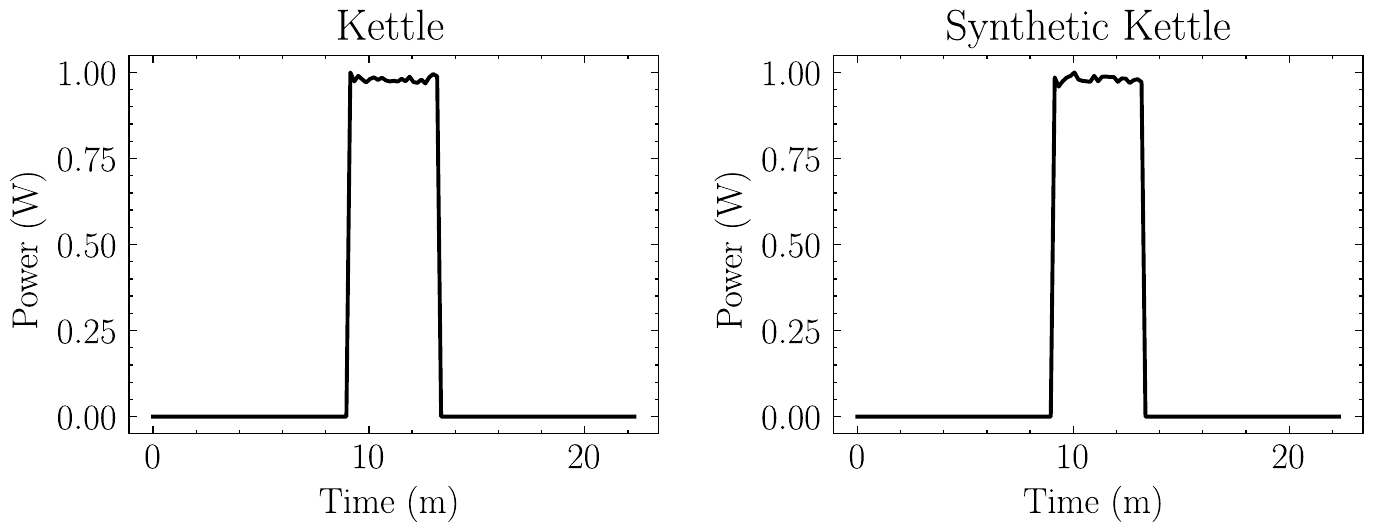}
  \caption{Primitive cycle of a kettle (left) and its corresponding parametrized model $p=p_1\cdot p_2\cdot p_4$ (right). The amplitude of both graphs is set to 1.}
  \label{fig:kettle}
\end{figure}

\begin{figure}[b!]
  \centering
  \includegraphics[width=.5\textwidth]{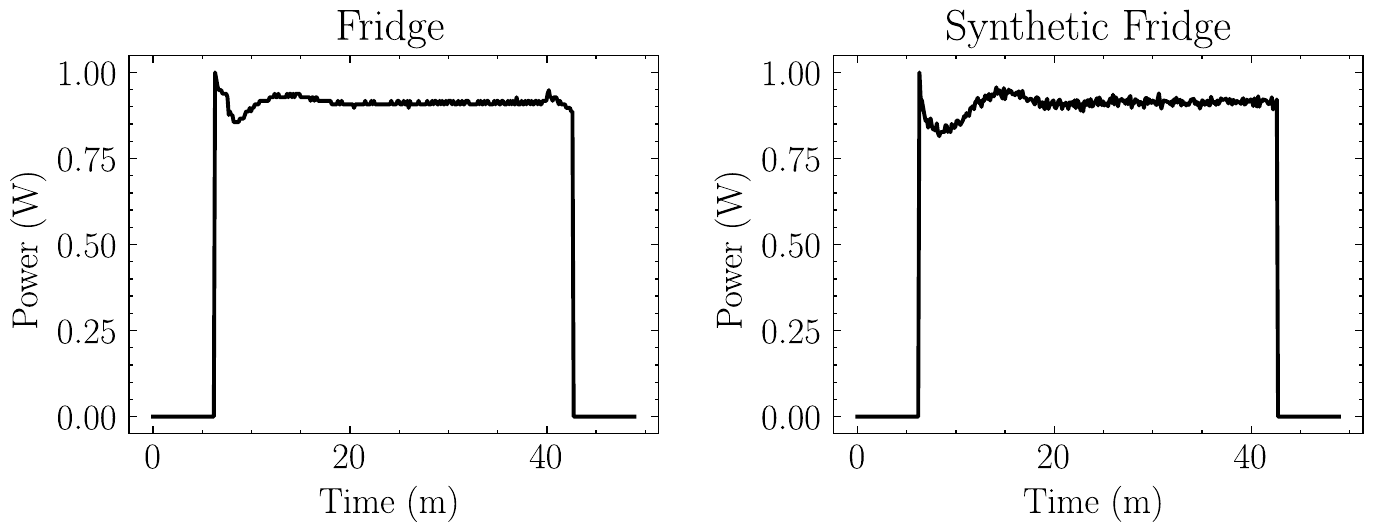}
  \caption{Primitive cycle of a fridge (left) and its corresponding parametrized model $p=p_1\cdot p_2\cdot p_3\cdot p_4$ (right). The amplitude of both graphs is set to 1.}
  \label{fig:fridge}
\end{figure}

\begin{figure}[b!]
  \centering
  \includegraphics[width=.5\textwidth]{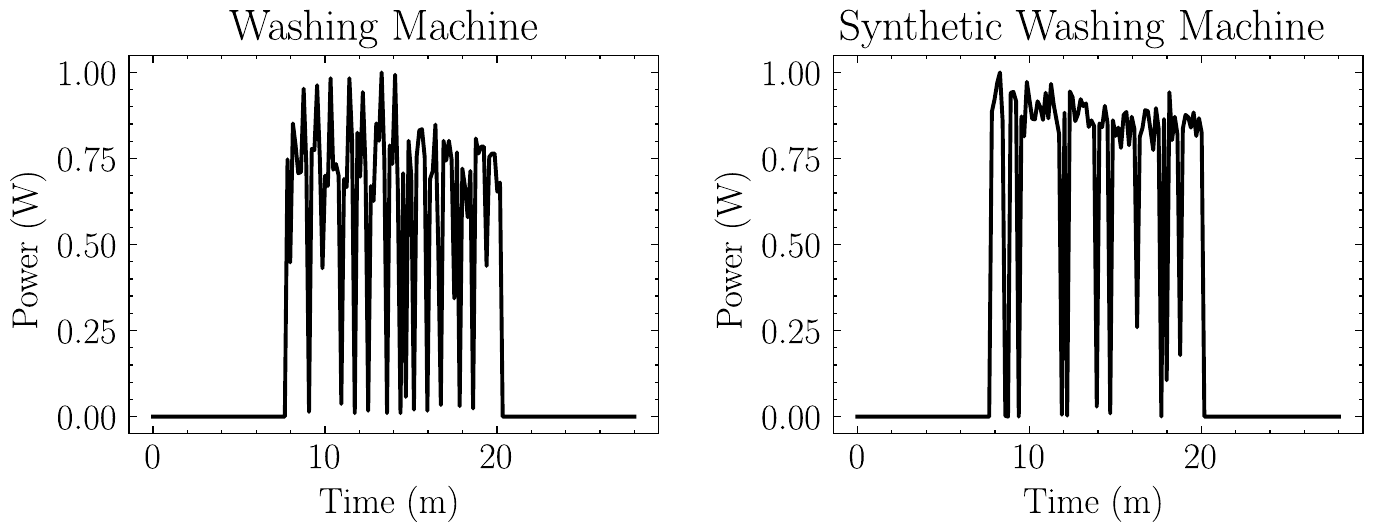}
  \caption{Primitive cycle of a washing machine (left) and its corresponding parametrized model $p=p_1\cdot p_2\cdot p_5$ (right). The amplitude of both graphs is set to 1.}
  \label{fig:wm}
\end{figure}

For example, the function $w=p_1\cdot p_2\cdot p_4$ can be related to most of the heating appliances e.g., water kettle, hair dryer, microwave etc. as in Fig.~\ref{fig:kettle}. The function $w=p_1\cdot p_2\cdot p_3\cdot p_4$ can approximate fridge cycles as in Fig.~\ref{fig:fridge}. The functions $w=p_1\cdot p_5$ and $w=p_1\cdot p_2\cdot p_5$ can represent the cycles of a washing machine and TV as in Fig.~\ref{fig:wm}.

\begin{figure}[t!]
  \centering
  \includegraphics[width=.5\textwidth]{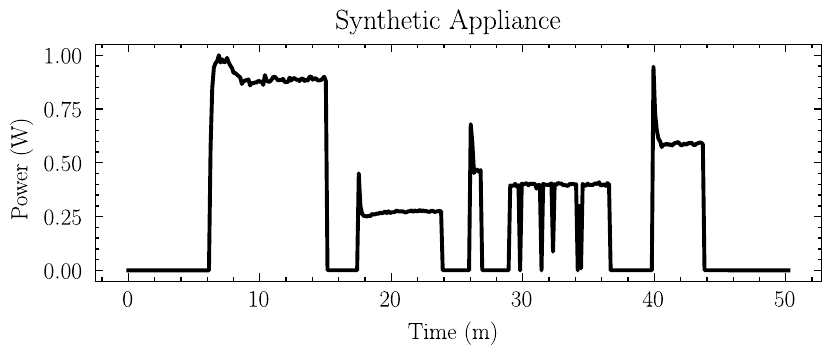}
  \caption{Synthetic appliance signature produced by the proposed approach and with random parametrization. The amplitude is set to 1.}
  \label{fig:rms}
\end{figure}

To generate a complete appliance signature (e.g. as in Fig.~\ref{fig:rms}), one can generate $n$ primitive cycles together with $n-1$ zero-consumption intervals by using the formulas:
\begin{equation}
    w_{i}=\prod p,
\end{equation}
\begin{equation}
\label{eq:seq1}
    W = \{\text{pad}(w_i;\Delta d_i)\}_{i=1}^{n-1}\bigcup w_n\}
\end{equation}
\begin{equation}
\label{eq:seq2}
    w' =
    \text{concat}(W).
\end{equation}

Based on the analysis of a UK-DALE dataset, the basis function $p_5$ takes place over the cycles which amplitude is much lower than the consecutive activations of other appliance regimes. Moreover, there are not so many appliances that contain cycles described by $p_5$. Thus, to achieve higher similarity between synthetic signatures and the real ones, we recommend bounding the resulting space of signatures by using the following conditioning: for each primitive cycle $i$, turn basis function $p_5$ into 1 with probability $P_b$ or if $a_i>\mathbb{E}[a]$.

To generate multiple appliances with multiple signatures, one can specify centroids for the parameters in Eqs.~\eqref{eq:p1},~\eqref{eq:p2},~\eqref{eq:p3},~\eqref{eq:p4},~\eqref{eq:p5} as done with the high sampling rate method. In addition, a few more parameters can be included in the centroids: the primitive cycle duration $\Delta t$, the delay in operation (i.e., the time span between two consecutive primitive cycles) $\Delta d$, the number of primitive cycles $n$. Thus, the centroid $Z$ of an appliance $k$ can be written as $Z_k=\{a_k, A_k, \tau_k, q_0, q_1, q_2, q_3, \alpha_k, \beta_k, \Delta t_k, \Delta d_k, n_k\}$. Note that all the model parameters are non-negative, which suggests that they may be sampled (except for the parameter $n$) from half-normal distributions with means corresponding to centroid coordinates and variance $\text{Var}_d$ that controls the diversity. Note that we defined parameter $n=n_k$ for all signatures of appliance $k$.

\section{Validation}
\label{sec:val}
To ensure that the synthetic data generated through our methods is suitable for energy disaggregation applications, we assessed the similarities between the distributions of synthetic and real data, and compared the proposed approach with our previous work \cite{10202926}. We used the Kullback-Leibler (KL) divergence as a measure to quantify the similarity between two distributions:
\begin{equation}
\label{eq:kl}
    D_{KL}(P \,||\, Q) = \sum_{x} P(x) \log\frac{P(x)}{Q(x)},
\end{equation}
where $P$ is a distribution of real data and $Q$ is a distribution of synthetic data; the lower the value of $D_{KL}$, the better $P$ is approximated by $Q$.

\setlength\tabcolsep{6.35pt}
\begin{table*}[t!]
\centering
\caption{Similarities between real and synthetic datasets estimated through Kullback-Leibler divergence $D_{KL}$.}
\label{tab:kl}
\begin{tabular}{@{}cccccccccccc@{}}
\toprule
\multicolumn{2}{c}{\textbf{Dataset}} & \multirow{2}{*}{\textbf{Sampling rate}}  & \multirow{2}{*}{\textbf{\# appliances}} & \multirow{2}{*}{\textbf{\# signatures}} & \multicolumn{6}{c}{\textbf{$D_{KL}$ over principal components}} & \multirow{2}{*}{$\overline{D}_{KL}$}\\
\cmidrule{1-2}\cmidrule{6-11}
Synthetic & Real &&  &    & 1             & 2             & 3    & 4             & 5    & 6    &               \\ \midrule\midrule[.1em]
Kamyshev et al. \cite{10202926} & \multirow{2}{*}{PLAID} & \multirow{2}{*}{high} & \multirow{2}{*}{16} &\multirow{4}{*}{1000} & 3.46    & 2.23    & 0.19    & 0.26    & 0.31    & 0.38   & 1.12                                 \\
Proposed  &  &  & &    & \textbf{0.63} & \textbf{1.26} & 0.36 & \textbf{0.25} & 0.57 & 1.04 & \textbf{0.69}\\
\cmidrule{1-4}\cmidrule{6-12}
Kamyshev et al. \cite{10202926} & \multirow{2}{*}{UK-DALE (house 1)} & \multirow{2}{*}{low} & \multirow{2}{*}{24} &  & 6.99    & 4.48    & 1.19    & 0.30    &  0.46   & 0.43   & 2.31                          \\
Proposed  &  & &  &   & \textbf{0.57} & \textbf{0.86} & \textbf{0.54} & 0.32 & 0.85 & \textbf{0.38} & \textbf{0.59}\\

\bottomrule
\end{tabular}
\end{table*}

We conducted two experiments, one for high sampling rate signatures and one for low sampling rate signatures. We extracted 1000 single-cycle waveforms of 16 appliances from PLAID for the first experiment, and 1000 signatures of 24 appliances from UK-DALE for the second. Note, that we padded and cropped UK-DALE's signatures which are below or above predefined duration, respectively. This step is needed in order to ensure that all signatures have identical duration for Principal Component Analysis (PCA).

Both datasets of real data were used to calculate the distributions $P$ from Eq.~\eqref{eq:kl}. To estimate the quality of the synthetic data produced by the proposed approach, we needed a baseline. We selected our previous work \cite{10202926} as such. Next, we generated two datasets with the same number of signatures and appliances as in the real dataset. We then applied PCA to the real and synthetic datasets, and reduced the dimensionality to 6 principal components. Six principal components explained 99\% of the variance of the original high sampling rate data. For the low sampling rate data, at least 60 principal components were needed. Further in the analysis, only first 6 components will be used.

To estimate the distributions $P$ and $Q$, we computed histograms out of 100 bins for each principal component. Finally, we calculated the pairwise KL-divergence (Eq.~\eqref{eq:kl}) between principal components of the real and synthetic datasets. The results for both experiments are summarized in Table~\ref{tab:kl}. 

\begin{figure}[b!]
  \centering
  \includegraphics[width=.5\textwidth]{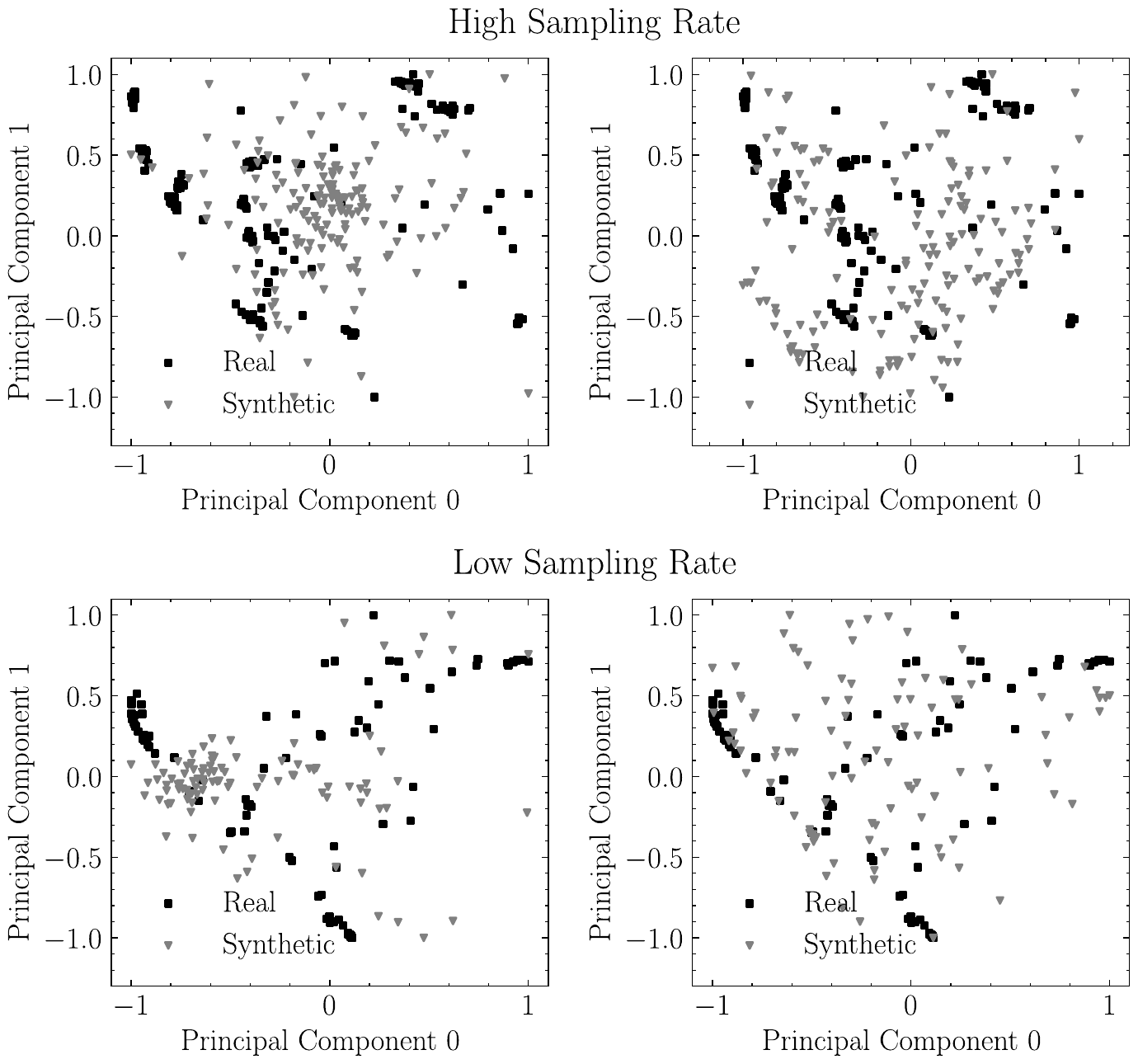}
  \caption{High and low sampling rate signatures of real and synthetic datasets projected onto 2D plane by using PCA. The top graphs represent the PLAID and synthetic datasets obtained by using method from \cite{10202926} (top left) and the proposed approach (top right). The bottom graphs show UK-DALE and synthetic datasets obtained by using the method from \cite{10202926} (bottom left) and our proposed approach (bottom right). The amplitude of graphs is set to 1.}
  \label{fig:scatter}
\end{figure}

As can be seen, the proposed approach is able to generate data that is significantly more similar to the real data than in previous works. That is, approximately 1.6 times for high sampling rate case and 3.9 times for low sampling rate case. Since principal components are ordered in descending order of their associated explained variance, the most important are the very first components. In this regard, our novel method is capable of producing signatures that are 5.5 and 12.3 times more similar to the original datasets for the first component, and 1.77 and 5.2 times for the second component.

Additionally, to show that synthetic signatures coincide with the real ones, we plotted two first principal components against each other for real and synthetic datasets (see Fig.~\ref{fig:scatter}). For better visualization purpose, we used only 100 arbitrary chosen signatures from each dataset. One can notice that the novel approach generates signatures whose 2D representations are scattered across a plane rather than concentrated around a specific point. This implies, that the signatures are diverse and not biased towards a particular waveform. This also demonstrates another benefit i.e., novel methods can potentially reconstruct a hypothetical manifold that describes all possible types of appliances and their states.

\section{Conclusion}
\label{sec:concl}
In this work, we proposed two novel physics-informed methods to generate appliance signatures at different sampling rates. Our methods have several advantages over previous works, including: (1) ability to generate diverse and unlimited variety of appliance signatures; (2) transparent and intuitive control over the underlying distributions; (3) no need in any input data for generating appliance signatures; (4) relatively simple mathematical model that makes the methods easy to reproduce and utilize. 

Through empirical validation using a KL-divergence and PCA, we have demonstrated that the synthetic data obtained by our methods is fair and utility-equivalent to real datasets, and potentially can generate all possible types of appliances and their states. The proposed methods are a valuable resource for researchers and engineers in the field of energy disaggregation. We believe that the mixture of real-world and synthetic datasets can significantly improve the performance of disaggregation algorithms, primarily due to its ability to mitigate data imbalance and data insufficiency challenges. The code for novel methods is available in the open source Python library Edframe: \href{https://github.com/arx7ti/edframe}{https://github.com/arx7ti/edframe}. 

\section{Acknowledgement}
This work was supported by the Skoltech program: Skolkovo Institute of Science and Technology -- Hamad Bin Khalifa University Joint Projects. 
Ilia Kamyshev would like to thank PhD student Dmitrii Kriukov for his valuable comments on Section~\ref{sec:val} of the paper.

\vspace{12pt}
\bibliographystyle{IEEEtran}
\bibliography{refs}

\end{document}